\documentstyle[preprint,aps]{revtex}
\preprint{USM-TH-162}
\begin{document}
\title{Brane Condensation and Confinement}
\author{P. Gaete$^{1}$\thanks{
E-mail: patricio.gaete@usm.cl} and C. Wotzasek $^{2}$\thanks{
E-mail: clovis@if.ufrj.br}}
\address{$^1$Departamento de F\'{\i}sica, Universidad T\'{e}cnica
F. Santa Mar\'{\i}a, Casilla 110-V, Valpara\'{\i}so, Chile. \\
$^2$Instituto de F\'{\i}sica, Universidade Federal do Rio de
Janeiro, 21945, Rio de Janeiro, Brazil.} \maketitle

\begin{abstract}
We study the static quantum potential for a theory of
anti-symmetric tensor fields that results from the condensation of
topological defects, within the framework of the gauge-invariant
but path-dependent variables formalism. Our calculations show that
the interaction energy is the sum of a Yukawa and a linear
potentials, leading to the confinement of static probe charges.
\end{abstract}
\smallskip

\section{INTRODUCTION}

One of the fundamental issues of theoretical physics
is that of the confinement for the fundamental constituents of
matter.
In fact distinction between the apparently related phenomena of
screening and confinement is of considerable importance in our
present understanding of gauge theories. Field theories that yield
the linear potential are important to particle physics, since
those theories may be used to describe the confinement of quarks
and gluons and be considered as effective theories of quantum
chromodynamics.

We study the confinement versus screening properties of some
theories of massless antisymmetric tensors, magnetically and
electrically coupled to topological defects that eventually
condense, as a consequence of the Julia--Toulouse mechanism
(JTM)\cite{JT}. This mechanism is the dual to the Higgs mechanism
and has been shown
to lead to a concrete massive antisymmetric theory with a jump of
rank. We show that in the presence of two tensor fields the
condensation induces not only a mass term and a jump of rank but
also a BF coupling which will be responsible for the change from
the screening to the confining phase of the theory.

An important issue here is the nature of the phase transition in
the presence of a finite condensate of topological defects. It is
this aspect, in $D=d+1$ dimensions for generic antisymmetric
tensors theories, that is of importance for us. This issue was
discussed long time ago
\cite{JT} in the framework of ordered solid-state media and more
recently in the relativistic context
\cite{QT}. The basic idea in Ref.\cite{JT} was to consider models
with non-trivial homotopy group able to support stable topological
defects characterized by a length scale $r=1/M$, where the mass
parameter $M$ is a cut-off for the low-energy effective field
theory.
The long wavelength fluctuations of the continuous distribution of
topological defects are the new hydrodynamical modes for the
effective theory that appear when topological defects condense. In
\cite{JT} there is an algorithm to identify these modes in the
framework of ordered solid-state media. However, due to the
presence of non-linear terms, the lack of relativistic invariance
and the need to introduce dissipation terms it becomes difficult
to write down an action for the phase with a condensate of
topological defects. In the relativistic context none of the above
problems is present.
In \cite{QT} an explicit form for the action in the finite
condensate phase, for generic compact antisymmetric field theories
was found. In this context the JTM is the natural generalization
of the confinement phase for a vector gauge field.

In this paper we make use of the JTM, as presented in \cite{QT},
to study the low-energy field theory of a pair of massless
anti-symmetric tensor fields, say $A_p$ and $B_q$ with $p+q+2=D$,
coupled electrically and magnetically to a large set of
$(q-1)$-branes, characterized by charge $e$ and a Chern-Kernel
$\Lambda_{p+1}$ \cite{HL}, that eventually condense. It is shown
that the effective theory that results displays the confinement
property by computing explicitly the effective potential for a
pair of static, very massive point probes.
Basically, we are interested in studying the JTM in model field
theories involving $B_q$ and $A_p$ coupled to a $(q-1)$-brane,
according to the following action
\begin{eqnarray} \label{R10}
{\cal S} &=&  \int \frac 12 \frac{(-1)^q}{(q+1)!}
\left[H_{q+1}(B_q) \right]^2 + e\, B_q J^{q}(\Lambda) +  \frac 12
\frac{(-1)^p}{(p+1)!}  \left[F_{p+1}(A_p) - e
\Lambda_{p+1}\right]^2,
\end{eqnarray}
and consider the condensation phenomenon when $\Lambda_{p+1}$
becomes the new massive mode of the effective theory. Our compact
notation here goes as follows. The field strength reads
$F_{p+1}\left(A_p\right)= F_{\mu_1 \mu_2 \ldots
\mu_{p+1}}=\partial_{[\mu_1}A_{\mu_2\cdots\mu_{p+1}]}$ and
$\Lambda_{p+1}=\Lambda_{\mu_1\cdots\mu_{p+1}}$ is a totally
anti-symmetric object of rank ($p+1$).
The conserved current $J^q(\Lambda)$ is given by a delta-function
over the world-volume of the ($q-1$)-brane \cite{Kleinert:kx}.
This conserved current may be rewritten in terms of the kernel
$\Lambda_{p+1}$ as
\begin{equation} J^q(\Lambda) = \frac
1{(p+1)!} \epsilon^{q,\alpha ,p+1}\partial_\alpha \Lambda_{p+1}\;
, \end{equation} and $\epsilon^{q,\alpha ,p+1} =
\epsilon^{\mu_1\ldots\mu_q,\alpha ,\nu_1\ldots\nu_{p+1}}$. This
notation will be used in the discussion of the JTM in the next
section as long as no chance of confusion occurs.


\section{The Julia--Toulouse Mechanism and the Action in the Condensed Phase}

Although the JT algorithm becomes problematic in the ordered solid
state media,
Quevedo and Trugenberg have shown that it leads to simple demands
in the study of compact antisymmetric tensor, where it produces
naturally the effective action for the new phase. They observed
that when the $(d-h-1)$-branes condense this generates a new scale
$\Delta$ related to the average density $\rho $ of intersection
points of the $(d-h)$-dimensional world-hypersurfaces of the
condensed branes with any $(h+1)$-dimensional hyperplane. The four
requirements
to describe effectively the dense phase are: (i) an action built
up to two derivatives in the new field possessing (ii) gauge
invariance, (iii) relativistic invariance  and, most important,
(iv) the need to recover the original model in the limit $\Delta
\to 0$.  One is therefore led to consider the action for the
condensate as
\begin{eqnarray} \label{RM15} {\cal S}_{\Omega}=\int
\frac {(-1)^h }{2 \Delta^2(h+1)!} \left[{F_{h+1}(\Omega_h)}
\right]^2 -\frac {(-1)^h\, h!}{2\, e^2} \left[\Omega_h - H_h
(\phi_{h-1})\right]^2,
\end{eqnarray}
where $H_{\mu_1\cdots\mu_h} =
\partial_{[\mu_1}\phi_{\mu_2\cdots\mu_{h}]}$ and
the underlying gauge invariance is manifest by the simultaneous
transformations $\Omega_{\mu_1\cdots\mu_h} \to
\Omega_{\mu_1\cdots\mu_h} +
\partial_{[\mu_1}\psi_{\mu_2\cdots\mu_h]}$ and
$\phi_{\mu_1\cdots\mu_{h-1}} \to \phi _{\mu_1\cdots\mu_{h-1}}+
\psi_{\mu_1\cdots\mu_{h-1}}$. Upon fixing this invariance one can
drop all considerations over $\phi_{h-1}$ after absorbing
$H_h(\phi_{h-1})$ into $\Omega_h$, so that the action describes
the exact number of degrees of freedom of a massive field whose
mass parameter reads $m=\Delta/e$. This process, named as JTM, is
dual to the Higgs mechanism. Here on the other hand, the new modes
generated by the condensation of magnetic topological defects
absorbs the original variables of the effective theory, thereby
acquiring a mass while in the Higgs mechanism it is the original
field that incorporates the degrees of freedom of the electric
condensate to acquire mass. This difference explains the change of
rank in the JT mechanism that is not present in the Higgs process.
In the limit $\Delta \to 0$ the only relevant field configurations
are those that satisfy the constraint $F_{h+1}(\Omega_h) =0$ whose
solution reads $\Omega_{\mu_1\cdots\mu_h} =
\partial_{[\mu_1}\psi_{\mu_2\cdots\mu_{h}]}$ where $\psi_{h-1}$ is an
$(h-1)$-anti-symmetric tensor field. The field $\psi_{h-1} $ can
then be absorbed into $\phi_{h-1} $ this way recovering the
original low-energy effective action before condensation.


The distinctive feature of the JT mechanism is that after
condensation $\Lambda_{p+1}$ is elevated to the condition of
propagating field. The new degree of freedom absorbs the degrees
of freedom of the tensor $A_p$ this way completing its
longitudinal sector. The new mode is therefore explicitly massive.
Since $A_p\to \Lambda_{p+1}$ there is a change of rank with
dramatic consequences. The last term in (\ref{R10}), displaying
the magnetic coupling between the field-tensor $F_{p+1}(A_p)$ and
the $(q-1)$-brane, becomes the mass term for the new effective
theory in terms of the tensor field $\Lambda_{p+1}$ and a new
dynamical term is induced by the condensation.
The minimal coupling of the $B_q$ tensor becomes responsible for
another contribution for the mass, this time of topological
nature. Indeed the second term (\ref{R10}) becomes a
``$B\wedge F(\Lambda)$" term between the remaining propagating
modes, inducing topological mass, in addition to the induced
condensed mass,
\begin{eqnarray}
\label{RM10} {\cal S}_{cond} &=& \int  \frac{(-1)^q}{2(q+1)!}
\left[H_{q+1}(B_q) \right]^2 + e\, B_q
\epsilon^{q,\alpha,p+1}\partial_\alpha \Lambda_{p+1} + \nonumber\\
&+& \int \frac {(-1)^{p+1}}{2 (p+2)!} \left[F_{p+2}(\Lambda_{p+1})
\right]^2 -\frac {(-1)^{p+1}\, (p+1)!}2 m^2 \Lambda_{p+1}^2.
\end{eqnarray}
Recall that the theory before condensation displayed two
independent fields coupled to a $(q-1)$--brane. The nature of the
two couplings were however different. The $A_p$ tensor, that was
magnetically coupled to the brane, was then absorbed by the
condensate after phase transition. On the other hand, the electric
coupling, displayed by the $B_q$ tensor, became a ``$B\wedge
F(\Lambda)$" topological term after condensation. There has been a
drastic change in the physical scenario.
We want next to obtain an effective action for the $B_q$ tensor.
To this end we shall next integrate out the field $\Lambda$
describing the condensate to obtain,
our final effective theory as
\begin{eqnarray}
\label{acaoeffetiva} {\cal S}_{eff} =\int \frac{(-1)^{q+1}}{2\,
(q+1)!} H_{q+1}(B_q) \left(1 +\frac{e^2}{\Delta^2 + m^2}
\right)H^{q+1}(B_q).
\end{eqnarray}

\section{Interaction Energy}

Next we examine the screening versus confinement issue. We shall
consider a specific example involving two Maxwell tensors coupled
electrically and magnetically to a point-charge such that after
condensation we end up with a Maxwell and a massive Kalb-Ramond
field (the condensate) coupled topologically to each other.
We shall calculate the interaction energy for the effective theory
between external probe sources by computing the expectation value
of the energy operator $H$ in the physical state $\left| \Phi
\right\rangle$ describing the sources, denote by $ \left\langle H
\right\rangle _\Phi$.
The Kalb-Ramond field $\Lambda_{\mu\nu}$ carrying the degrees of
freedom of the condensate is integrate out leading to
\begin{equation}
{\cal L} =  - \frac{1}{4}F_{\mu \nu } \left( {1 + \frac{{e^2
}}{{\triangle^2 + m^2 }}} \right)F^{\mu \nu }  - A_0 J^0 ,
\label{KR15}
\end{equation}
where
$J^0$ is an external current.
We observe
that the limits $e\to 0$ or $m\to 0$ are well defined and lead
to a pure Maxwell theory or to a (topologically) massive model.
Since the probe charges only couple to the Maxwell fields, the
Kalb- Ramond condensate will not contribute to their interaction
energy in the first case because in the limit where the parameter
$e\to 0$ the Maxwell field and the condensate decouple. The second
limit means that we are back to the non-condensed phase. As so the
confinement of the probe charges are expected to disappear being
taken over by an screening phase controlled by the parameter $e$
playing the role of topological mass.

Once this is done, the canonical quantization of this theory from
the Hamiltonian point of view follows straightforwardly. The
canonical momenta read $\Pi ^\mu   = - \left( {1 + \frac{{e^2
}}{{\Delta ^2 + m^2 }}} \right)F^{0\mu }$ with the only
nonvanishing canonical Poisson brackets being $\left\{ {A_\mu
\left( {t,x} \right),\Pi ^\nu  \left( {t,y} \right)} \right\} =
\delta _\mu ^\nu  \delta \left( {x - y} \right)$. Since $\Pi_0$
vanishes we have the usual primary constraint $\Pi_0=0$, and $\Pi
^i  = \left( {1 + \frac{{e^2 }}{{\Delta ^2 + m^2 }}}
\right)F^{i0}$. The canonical Hamiltonian is thus
\begin{equation}
H_C  = \int {d^3 } x\left\{ { - \frac{1}{2}\Pi ^i \left( {1 +
\frac{{e^2 }}{{\Delta ^2  + m^2 }}} \right)^{ - 1} \Pi _i  + \Pi
^i \partial _i A_0  + \frac{1}{4}F_{ij} \left( {1 + \frac{{e^2
}}{{\Delta ^2  + m^2 }}} \right)F^{ij}  + A_0 J^0 }.
\right\}.\label{KR25}
\end{equation}
Time conservation of the primary constraint $ \Pi _0$ leads to the
secondary Gauss-law constraint $\Gamma _1 \left( x \right) \equiv
\partial _i \Pi ^i - J^0 = 0$. The preservation of $\Gamma_1$ for
all times does not give rise to any further constraints. The
theory is thus seen to possess only two constraints, which are
first class, therefore the theory described by $(\ref{KR15})$ is a
gauge-invariant one. The extended Hamiltonian that generates
translations in time then reads $H = H_C  + \int {d^3 } x\left(
{c_0 \left( x \right)\Pi _0 \left( x \right) + c_1 \left( x
\right)\Gamma _1 \left( x \right)} \right)$, where $c_0 \left( x
\right)$ and $c_1 \left( x \right)$ are the Lagrange multiplier
fields. Moreover, it is straightforward to see that $\dot{A}_0
\left( x \right)= \left[ {A_0 \left( x \right),H} \right] = c_0
\left( x \right)$, which is an arbitrary function. Since $ \Pi^0 =
0$ always, neither $ A^0 $ nor $ \Pi^0 $ are of interest in
describing the system and may be discarded from the theory.

The quantization of the theory requires the removal of nonphysical
variables, which is done by imposing a gauge condition such that
the full set of constraints becomes second class. A convenient
choice is found to be \cite{Pato} $\Gamma _2 \left( x \right)
\equiv \int\limits_{C_{\xi x} } {dz^\nu } A_\nu \left( z \right)
\equiv \int\limits_0^1 {d\lambda x^i } A_i \left( {\lambda x}
\right) = 0$, where  $\lambda$ $(0\leq \lambda\leq1)$ is the
parameter describing the spacelike straight path $ x^i = \xi ^i  +
\lambda \left( {x - \xi } \right)^i $, and $ \xi $ is a fixed
point (reference point). There is no essential loss of generality
if we restrict our considerations to $ \xi ^i=0 $. In this case,
the only nonvanishing equal-time Dirac bracket is
\begin{equation}
\left\{ {A_i \left( x \right),\Pi ^j \left( y \right)} \right\}^ *
=\delta{ _i^j} \delta ^{\left( 3 \right)} \left( {x - y} \right) -
\partial _i^x \int\limits_0^1 {d\lambda x^j } \delta ^{\left( 3
\right)} \left( {\lambda x - y} \right). \label{KR45}
\end{equation}

We now turn to the problem of obtaining the interaction energy
between pointlike sources in the model under consideration. The
state $\left| \Phi \right\rangle$ representing the sources is
obtained by operating over the vacuum with creation/annihilation
operators. We want to stress that, by construction, such states
are gauge invariant. In the case at hand we consider the
gauge-invariant stringy $\left|{\overline \Psi  \left( \bf y
\right)\Psi \left( {\bf y^ \prime } \right)} \right\rangle$, where
a fermion is localized at ${\bf y}\prime$ and an antifermion at $
{\bf y}$ as follows \cite{Dirac2},
\begin{equation}
\left| \Phi  \right\rangle  \equiv \left| {\overline \Psi  \left(
\bf y \right)\Psi \left( {\bf y}\prime \right)} \right\rangle  =
\overline \psi \left( \bf y \right)\exp \left(
{iq\int\limits_{{\bf y}\prime}^{\bf y} {dz^i } A_i \left( z
\right)} \right)\psi \left({\bf y}\prime \right)\left| 0
\right\rangle, \label{KR60}
\end{equation}
where $\left| 0 \right\rangle$ is the physical vacuum state and
the line integral appearing in the above expression is along a
spacelike path starting at ${\bf y}\prime$ and ending $\bf y$, on
a fixed time slice. It is worth noting here that the strings
between fermions have been introduced in order to have a
gauge-invariant function $\left| \Phi  \right\rangle $. In other
terms, each of these states represents a fermion-antifermion pair
surrounded by a cloud of gauge fields sufficient to maintain gauge
invariance. As we have already indicated, the fermions are taken
to be infinitely massive (static).

From our above discussion, we see that $\left\langle H
\right\rangle _\Phi$ reads
\begin{equation}
\left\langle H \right\rangle _\Phi   = \left\langle \Phi
\right|\int {d^3 } \left\{ { - \frac{1}{2}\Pi _i \left( {1 -
\frac{{e^2 }}{{\nabla ^2  - m^2 }}} \right)^{ - 1} \Pi ^i }
\right\}\left| \Phi  \right\rangle, \label{KR65}
\end{equation}
where, in this static case, $\Delta ^2 = - \nabla ^2$. Observe
that when $e=0$ we obtain the pure Maxwell theory, as mentioned
after (\ref{KR15}).  From now on we will suppose $e\neq 0$.

Next, from the foregoing Hamiltonian analysis, $\left\langle H
\right\rangle _\Phi$ becomes $\left\langle H \right\rangle _\Phi =
\left\langle H \right\rangle _0  + V^{\left( 1 \right)}  +
V^{\left( 2 \right)}$, where $\left\langle H \right\rangle _0  =
\left\langle 0 \right|H\left| 0 \right\rangle$. The $V^{\left( 1
\right)}$ and $V^{\left( 2 \right)}$ terms are given by:
\begin{equation}
V^{\left( 1 \right)}  =  - \frac{{q^2 }}{2}\int {d^3 x} \int_{\bf
y}^{\bf y^\prime} {dz^\prime_i } \delta ^{\left( 3 \right)} \left(
{x - z^\prime} \right)\frac{1}{{\nabla _x^2  - M^2 }}\nabla _x^2
\int_{\bf y}^{\bf y^\prime} {dz^i } \delta ^{\left( 3 \right)}
\left( {x - z} \right), \label{KR80}
\end{equation}
and
\begin{equation}
V^{\left( 2 \right)}  =   \frac{{q^2 m^2}}{2}\int {d^3 x}
\int_{\bf y}^{\bf y^\prime} {dz^\prime_i } \delta ^{\left( 3
\right)} \left( {x - z^\prime} \right)\frac{1}{{\nabla _x^2  - M^2
}} \int_{\bf y}^{\bf y^\prime} {dz^i } \delta ^{\left( 3 \right)}
\left( {x - z} \right), \label{KR85}
\end{equation}
where $ M^2\equiv{m^2+ e^2} $ and the integrals over $z^i$ and
$z^\prime_i$ are zero except on the contour of integration.

The $V^{\left( 1 \right)}$ term may look peculiar, but it is
nothing but the familiar Yukawa interaction plus self-energy
terms. In effect, as was explained in Ref. \cite{GG2}, the
expression (\ref{KR80}) can also be written as
\begin{equation}
V^{\left( 1 \right)}  = \frac{{e^2 }}{2}\int_{\bf y}^{{\bf
y}^{\prime}  } {dz_i^{\prime}}\partial _i^{z^{\prime}} \int_{\bf
y}^{{\bf y}^{\prime}} {dz^i }\partial _z^i G\left( {{\bf
z}^{\prime},{\bf z}} \right)=   - \frac{{q^2 }}{{4\pi }}\frac{{e^{
- M|{\bf y} - {\bf y}^ {\prime}| } }}{{|{\bf y} - {\bf
y}^{\prime}|}}, \label{KR90}
\end{equation}
where we used that the Green function $G({\bf z}^{\prime},{\bf z})
= \frac{1}{{4\pi }}\frac{{e^{ - M|{\bf z}^{\prime} - {\bf z}|}
}}{{|{\bf z}^{\prime} - {\bf z}|}}$ and remembered that the
integrals over $z^i$ and $z_i^{\prime}$ are zero except on the
contour of integration. The expression
then reduces to the Yukawa-type potential after subtracting the
self-energy terms.

We now turn our attention to the calculation of the $V^{\left( 2
\right)}$ term, which is given by
\begin{equation}
V^{\left( 2 \right)}  = \frac{{q^2 m^2 }}{2}\int_{\bf y}^{{\bf
y}^{\prime}  } {dz^{{\prime} i} } \int_{\bf y}^{{\bf y}^{\prime} }
{dz^i } G({\bf z}^{\prime} ,{\bf z}). \label{KR105}
\end{equation}
It is appropriate to observe here that the above term is similar
to the one found for the system consisting of a gauge field
interacting with a massive axion field \cite{GG2}.
Notwithstanding, in order to put our discussion into context it is
useful to summarize the relevant aspects of the calculation
described previously \cite{GG2}. In effect, as was explained in
Ref. \cite{GG2}, by using the Green function in momentum space,
that is, $\frac{1}{{4\pi }}\frac{{e^{ - M|{\bf z}^{\prime}   -
{\bf z}|} }}{{|{\bf z}^{\prime} - {\bf z}|}} = \int {\frac{{d^3
k}}{{\left( {2\pi } \right)^3 }}\frac{{e^{i{\bf k} \cdot \left(
{{\bf z}^{\prime}- {\bf z}} \right)} }}{{{\bf k}^2  + M^2 }}}$,
the expression (\ref{KR105}) can also be written as
\begin{equation}
V^{\left( 2 \right)}   = q^2 m^2 \int {\frac{{d^3 k}}{{\left(
{2\pi } \right)^3 }}} \left[ {1 - \cos \left( {{\bf k} \cdot {\bf
r}} \right)} \right]\frac{1}{{({\bf k}^2  + M^2)
}}\frac{1}{{\left( {{\bf {\hat n}} \cdot {\bf k}} \right)^2 }},
\label{KR115}
\end{equation}
where ${\bf {\hat n}}\equiv \frac{{{\bf y} - {\bf
y}^{\prime}}}{{|{\bf y} - {\bf y}^{\prime}| }}$ is a unit vector
and ${\bf r}={\bf y}-{\bf y^{\prime}}$ is the relative vector
between the quark and antiquark. Since ${\bf {\hat n}}$ and ${\bf
r}$ are parallel, we get accordingly $V^{\left( 2 \right)} =
\frac{{q^2 m^2 }}{{8\pi ^3 }}\int\limits_{ - \infty }^\infty
{\frac{{dk_r }}{{k_r^2 }}} \left[ {1 - \cos \left( {k_r r}
\right)} \right]\int\limits_0^\infty  {d^2 k_T \frac{1}{{(k_r^2 +
k_T^2  + M^2) }}}$, where $k_T$ denotes the momentum component
perpendicular to ${\bf r}$. Integration over $k_T$ yields $
V^{\left( 2 \right)}  = \frac{{q^2 m^2 }}{{8\pi ^2 }}\int\limits_{
- \infty }^\infty  {\frac{{dk_r }}{{k_r^2 }}} \left[ {1 - \cos
\left( {k_r r} \right)} \right]\ln \left( {1 + \frac{{\Lambda ^2
}}{{k_r^2  + M^2 }}} \right)$, where $\Lambda$ is an ultraviolet
cutoff. We also observe at this stage that similar integral was
obtained independently in Ref.\cite{Suganuma} in the context of
the dual Ginzburg-Landau theory by an entirely different approach.

We now proceed to compute the previous integral. For this purpose
we introduce a new auxiliary parameter $\varepsilon$ by making in
the denominator of the previous integral the substitution
$k_r^2\rightarrow k_r^2+\varepsilon^2$. Thus it follows that $
V^{\left( 2 \right)} \equiv \lim _ {\varepsilon  \to 0}
{\widetilde V}^{\left( 2 \right)}= \lim _{\varepsilon \to
0}\frac{{q^2 m^2 }}{{8\pi ^2 }}\int\limits_{ - \infty }^\infty
{\frac{{dk_r }}{{(k_r^2  + \varepsilon ^2) }}} \left[ {1 - \cos
\left( {k_r r} \right)} \right]\ln \left( {1 + \frac{{\Lambda ^2
}}{{k_r^2  + M^2 }}} \right)$. We further note that the
integration on the $k_r$-complex plane yields ${\widetilde
V}^{\left( 2 \right)} = \frac{{q^2 m^2 }}{{8\pi }}\left( {\frac{{1
- e^{ - \varepsilon |{\bf y} - {{\bf y}^\prime}|  } }}{\varepsilon
}} \right)\ln \left( {1 + \frac{{\Lambda ^2 }}{{M^2 - \varepsilon
^2 }}} \right)$. Taking the limit $\varepsilon  \to 0$, this
expression then becomes $ V^{\left( 2 \right)}  = \frac{{q^2 m^2
}}{{8\pi }}|{\bf y} - {{\bf y}^\prime}| \ln \left( {1 +
\frac{{\Lambda ^2 }}{{M^2 }}} \right)$.

 This, together with Eq.(\ref{KR90}), immediately shows
that the potential for two opposite charges located at ${\bf y}$
and ${\bf y^\prime}$ is given by
\begin{equation}
V(L) =  - \frac{{q^2 }}{{4\pi }}\frac{{e^{ - ML} }}{L} +
\frac{{q^2 m^2 }}{{8\pi }}L\ln \left( {1 + \frac{{\Lambda ^2
}}{{M^2 }}} \right), \label{KR145}
\end{equation}
where $L\equiv|{\bf y}-{\bf {y^\prime}}|$.

\section{Final Remarks}

We have studied the confinement versus screening issue for a pair
of antisymmetric tensors coupled to topological defects that
eventually condense, giving a specific realization of the
Julia--Toulouse phenomenon. We have seen that the Julia--Toulouse
mechanism for a couple of massless antisymmetric tensors is
responsible for the appearance of mass and the jump of rank in the
magnetic sector while the electric sector becomes a BF--type
coupling. The condensate absorbs and replaces one of the tensors
and becomes the new massive propagating mode but does not couple
directly to the probe charges.  The effects of the condensation
are however felt through the BF coupling with the remaining
massless tensor.  It is therefore not surprising that they become
manifest in the interaction energy for the effective theory.  We
have obtained the effective theory for the condensed phase in
general and computed the interaction energy between two static
probe charges, in a specific example, in order to test the
confinement versus screening properties of the effective model.
Our results show that the interaction energy in fact contains a
linear confining term and an Yukawa type potential. It can be
observed that confinement completely disappears in the limit $m\to
0$ while the screening takes over controlled by the topological
mass parameter instead. Although we have considered the case where
the effective model consists of the BF--coupling between a
Kalb-Ramond field (that represents the condensate) and a Maxwell
field, our results seem to be quite general.  A direct calculation
for tensors of arbitrary rank in the present approach is however a
quite challenging problem that we hope to be able to report in the
future.

\end{document}